\documentstyle{ioplppt}                
\begin{document}
\title{Temperature effect in the Casimir attraction of a thin metal film}

\author{V N Dubrava and V A Yampol'skii
\ftnote{1}{E-mail: yam@ire.kharkov.ua}}

\address{Institute for Radiophysics and Electronics, National Academy of
Sciences of Ukraine,\\ 12 Academician Proskura Str., Kharkov 310085,
Ukraine}

\begin{abstract}
The Casimir effect for conductors at arbitrary temperatures is
theoretically studied. By using the analytical properties of the Green
functions and applying the Abel-Plan formula to Lifshitz's equation, the
Casimir force is presented as sum of a temperature dependent and vacuum
contributions of the fluctuating electromagnetic field. The general results
are applied to the system consisting of a bulk conductor and a thin metal
film. It is shown that a characteristic frequency of the thermal
fluctuations in this system is proportional to the square root of a
thickness of the metal film. For the case of the sufficiently high
temperatures when the thermal fluctuations play the main role in the Casimir
interaction, this leads to the growth of the effective dielectric
permittivity of the film and to a disappearance of the dependence of
Casimir's force on the sample thickness.

\end{abstract}



\section{Introduction}
The theoretical (see, for example, Refs.~\cite{1,2,3,5,5*,5**,4,6,7}) and
experimental~\cite{8,9,10,11} study of the Casimir effect has more than
fifty years of history. The kernel of this phenomenon consists in the
fluctuation electromagnetic interaction of uncharged bodies. For metals
with a high value of the conductivity $\sigma$, the Casimir interaction
manifests itself as the attractive force $f_0$ which varies as the inverse
fourth power of the distance $a$ between the plates~\cite{1},
\begin{equation}
f_{0}=-\frac{\pi^2}{240}\frac{\hbar c}{a^4} \qquad (T=0,\,\,\sigma\to\infty),
\label{I1}
\end{equation}
where $c$ is the speed of light and $T$ is the temperature.

With an increase of the temperature, but under the condition
\begin{equation}
\frac{kT}{\hbar} \ll \frac{c}{a},
\label{I2}
\end{equation}
an additional term proportional to the fourth power of the
temperature appears in the Casimir force (see Ref.~\cite{12} and references
therein),
\begin{equation}
\Delta f(T)=-\frac{\pi^2}{45}\frac{(kT)^4}{(\hbar c)^3}.
\label{I3}
\end{equation}
This term was derived under an assumption that the thermal equilibrium
between the matter and the radiation takes place. The additional
$a$-independent attractive force between the metal plates arises as a result
of the pressure of the thermal radiation outside of the plates.  Under the
inverse inequality,
\begin{equation}
\frac{kT}{\hbar} \gg \frac{c}{a},
\label{I4}
\end{equation}
the Casimir force is completely defined by the temperature and is described
by the formula
\begin{equation}
f(T)=-1.2\frac{kT}{8\pi a^3}
\label{I5}
\end{equation}
with the exponential accuracy. At $T=300$~K, the parameter $\hbar c/kT$ is
about $30\,\,\mu {\rm m}$. So, the Casimir force between the bulk metal
plates displays a weak temperature dependence (Eq.~(\ref{I3})) in the range
of the realistic separations $a$ about $0.1-1\,\,\mu {\rm m}$.

The temperature effects in the Casimir force could be in the forefront if
the interacting objects are thin metal films. Indeed, formula (\ref{I1}) is
obtained under an assumption that the thickness $d$ of the plates is the
greatest parameter with the dimension of the length. As was shown in
Refs.~\cite{13,14}, the asymptotical formula (\ref{I1}) appears to be invalid
for the thin metal plates if the inequality
\begin{equation}
\omega_p\sqrt{\frac{d}{a}}=\omega_c\ll\omega_p ,\frac{c}{a}
\label{I6}
\end{equation}
is fulfilled. Here $\omega_p$ is the plasma frequency, $\omega_c$ is the
characteristic frequency of the fluctuating electromagnetic field. In this
case the collective properties of the electron subsystem of the metal are
important in forming the Casimir force. Specifically, the evaluation
\begin{equation}
f_0\propto -\frac{\hbar\omega_c^2}{\nu +\omega_c}\frac{1}{a^3}
\quad (T=0),
\label{I7}
\end{equation}
is valid if the metal film of a thickness $d$ with a weak reflecting power
interacts with the bulk metal ($\nu$ is the frequency of the electron bulk
collisions). The decrease of the absolute value of the Casimir force
$f(T=0)$ allows one to emphasize the temperature dependence $f(T)$ even in
the range of the realistic separations $a\sim 0.1-1\,\,\mu {\rm m}$. The
study of this dependence is a subject of the present paper.

\section{Statement of the problem. The basic equations}
The general formula for the force of Casimir interaction between dielectric
slabs with arbitrary dielectric constants $\epsilon$ was originally derived
by E.~Lifshitz~\cite{2} (see, also, Refs.~\cite{3,4,15,16}). The Casimir
force is presented in this formula as a functional defined on the set
of functions $\epsilon ({\rm i}\omega_n)$ of a discrete variable $\omega_n
=2\pi nkT$ ($n=0,1,2, \dots$). We use Lifshitz's formula for the system
comprising a bulk conductor and a thin metal film of thickness $d$ separated
by a distance $a$. The system of coordinates is chosen so that the $x$-axis
is perpendicular to the plane of interacting plates. The conductivity of our
system as the function of the coordinate $x$ is
\begin{equation}
\fl
\sigma (x)=\sigma[\theta (-x+a+d)-\theta (-x+a)]+\sigma_\infty\theta (x)
=\left\{
\begin{array}{ll}
0, &\,\,\, x\in S_I,\,\, x\in S_{III},\\
\sigma, &\,\,\, x\in S_{II},\\
\sigma_\infty , &\,\,\, x\in S_{IV},
\end{array}
\right.\nonumber
\label{s}
\end{equation}
where $\theta (x)$ is the Heaviside function; $S_I =(-\infty ,-a-d)$ and
$S_{III}=(-a,0)$ are vacuum domains; $S_{II}=(-a-d,-a)$ and
$S_{IV}=(0,\infty)$ are regions occupied by the metal film and the bulk
conductor, respectively. We do not make specific supposition about
the conductivity $\sigma_\infty$ of the bulk conductor because its value
does not affect the final result. As for the conductivity of the metal film,
we take it in the $\tau$-approximation with the following frequency
dependence:
\begin{equation}
\sigma (\omega)=\frac{\omega_p^2}{\nu - {\rm i}\omega}.
\label{sigma}
\end{equation}

Lifshitz's formula for the Casimir force can be expressed in terms of the
conductivity (\ref{s}) of our system,
\begin{equation}
F[\sigma]=-kT{\sum_{n=0}^\infty}^\prime\int {\rm d}\vec r\,
\frac{\delta\sigma^{(M)}(x|\omega_n)}{\delta a}
\Gamma^{(M)}_{ii}(\vec r,\vec r\,|\omega_n),
\label{1}
\end{equation}
where Matsubar's conductivity $\sigma^{(M)}(\omega_n)$ is related to the
frequency dispersion of the metal conductivity $\sigma(\omega)$ by the
relation,
\begin{equation}
\sigma^{(M)}(\omega_n)=\sigma({\rm i}\omega_n);
\label{2}
\end{equation}
$\Gamma^{(M)}$ is the temperature Green function of the electromagnetic
field; the prime on the sum symbol indicates that the term with $n=0$
is taken with half the weight. By using the analytical properties of the
Green functions and the Abel-Plan formula for summing up the series,
Eq.~(\ref{1}) can be rewritten in the integral form (see Appendix A),
\begin{eqnarray}
F[\sigma] &=-\frac{1}{2\pi}\int {\rm d}\vec r\,
\left\{\int\limits_0^{\infty}{\rm d}\zeta
\frac{\delta\sigma(x|{\rm i}\zeta)}{\delta a}
\Gamma_{ii}(\vec r,\vec r\,|{\rm i}\zeta) \right. \nonumber\\
&\left.+2\int\limits_{0}^\infty{\rm d}\omega {\rm Im}
\left[\frac{\delta\sigma(x|\omega)}{\delta a}
\Gamma_{ii}(\vec r,\vec r\,|\omega)\right]
\left(e^{\hbar\omega/kT}-1\right)^{-1}\right\}.
\label{A4}
\end{eqnarray}

The first term in Eq.~(\ref{A4}) describes the Casimir force at zeroth
temperature and is obtained from Eq.~(\ref{1}) through the simple change of
the summation by integrating over the imaginary frequency. The second term
provides the temperature-dependent contribution to the Casimir force which
is suppressed by the small exponential factor $\exp(-\hbar\omega /kT)$ at
$T\to 0$. Contrary to the low-temperature case, this term can be governing
in the force (\ref{A4}) at sufficiently high temperatures.

In order to simplify the general formula~(\ref{1}), we introduce the
transverse spatial Fourier transformation
\[
\Gamma^{(M)}_{ik} (\vec r,{\vec r\,}^\prime |\omega_n)=\int\frac{{\rm d}
\vec q}{(2\pi)^2}\exp [{\rm i}\vec q \,(\vec r - {\vec r\,}^\prime)_\perp]
\Gamma^{(M)}_{ik} (x,x^\prime |q^2,\omega_n).
\]
At infinitesimal displacement $\delta a$ of the bulk conductor, the
conductivity $\sigma (x)$ changes by
\[
\delta\sigma (x)=-\sigma_\infty\delta (x) \delta a.
\]
Therefore, formula (\ref{1}) for the Casimir force can be rewritten in the
final form,
\begin{equation}
\left.f=\frac{F}{A}=kT\int\limits_0^\infty\frac{{\rm d}q^2}
{4\pi}{\sum_{n=0}^\infty}^\prime\sigma_\infty \Gamma^{(M)}_{ii}
(x,x^\prime |q^2 ,\omega_n)\right|_{x \to 0, \, x^\prime \to 0},
\label{3}
\end{equation}
where $A$ is the area of slabs. We interpret the limiting process in
Eq.~(\ref{3}) as one in which $x$ and $x^\prime$ tend to the interface
from opposite sides, $x\to -0$ and $x^\prime\to +0$. The formula (\ref{3})
defines the force acting on the unit area of the bulk conductor from the
metal film. The positive force corresponds to the repulsion of bodies and the
negative one to the attraction.

In terms of "transverse electric" and "transverse magnetic" Green's
functions $g^e$ and $g^m$~\cite{3}, we have
\[
\fl
\lim_{x,x^\prime \to 0}\Gamma^{(M)}_{ii} (x,x^\prime)=\lim_{x,x^\prime \to 0}
\left[\omega_n g^e (x,x^\prime)-\omega_n^{-1}\left(\partial_x\frac{1}
{\epsilon_\infty}
{\partial_x}^\prime +\frac{q^2}{\epsilon_\infty}\right)g^m
(x,x^\prime)\right],
\]
where $g^e$ and $g^m$ are defined by
\begin{equation}
[-\partial_x^2 + q^2 + \omega_n^2\epsilon(x|\omega_n)]g^e (x,x^\prime) =
\delta(x - x^\prime)
\label{4}
\end{equation}
and
\begin{equation}
\left[-\partial_x\frac{1}{\epsilon(x|\omega_n)}\partial_x +
\frac{q^2}{\epsilon(x|\omega_n)} + \omega_n^2 \right]g^m (x,x^{\prime}) =
\delta(x - x^\prime).
\label{5}
\end{equation}
Here
\begin{equation}
\epsilon(x|\omega_n) =1+\frac{\sigma (x|{\rm i}\omega_n)}{\omega_n},
\label{epsil}
\end{equation}
is the effective dielectric permittivity of a metal taken at the imaginary
frequency.

Thus, to analyze the temperature dependence of the Casimir force we should
solve the set of Eqs.~(\ref{4}), (\ref{5}) and substitute the obtained
function $\Gamma^{(M)}_{ii} (x=0,x^\prime=0|q^2 ,\omega_n)$ into
Eq.~(\ref{3}).

\section{Temperature dependence of the Casimir force}
While solving the set of Eqs.~(\ref{4}) and (\ref{5}), we are interested in
Green's functions with the argument $x^\prime$ within the region $S_{IV}$
occupied by the bulk conductor. For $x^\prime \in S_{IV}$, the general
solutions of Eqs.~(\ref{4}) and (\ref{5}) have the following form:
\begin{eqnarray}
g_I^{(e,m)}     &=Ae^{kx-\kappa_\infty x^\prime},\quad
g_{II}^{(e,m)}=\left(B_1 e^{\kappa x}+B_2 e^{-\kappa x}\right)
e^{-\kappa_\infty x^\prime};\nonumber\\
g_{III}^{(e,m)} &=\left(C_1 e^{kx}+C_2 e^{-kx}\right)
e^{-\kappa_\infty x^\prime}; \nonumber\\
g_{IV}^e  &=(2\kappa_\infty)^{-1}\left(e^{-\kappa_\infty |x-x^\prime|}+
re^{-\kappa_\infty (x+x^\prime )}\right); \nonumber\\
g_{IV}^m  &=\epsilon_\infty (2\kappa_\infty)^{-1}
\left(e^{-\kappa_\infty |x-x^\prime|}+re^{-\kappa_\infty (x+x^\prime )}
\right),
\nonumber
\end{eqnarray}
where
\[
k=\sqrt{q^2+\omega_n^2},\quad \kappa_\infty
=\sqrt{q^2+\epsilon_\infty\omega_n^2},
\quad \kappa =\sqrt{q^2+\epsilon\omega_n^2}.
\]

Determining constants $A$, $B_{1,2}$, $C_{1,2}$, and $r$ from the boundary
conditions to Eqs.~(\ref{4}) and (\ref{5}), we obtain for the difference
\begin{equation}
\Gamma_{ii}(a) - \Gamma_{ii}(a\to\infty)\equiv {\rm reg} \Gamma_{ii}(a)
\label{6}
\end{equation}
the expression
\begin{equation}
{\rm reg}\Gamma_{ii}(a)=-k\sigma_\infty^{-1}
\frac{2\epsilon kd\exp(-2ka)}{2+\epsilon kd[1-\exp(-2ka)]}+\cdots .
\label{7}
\end{equation}
This formula is derived under inequalities (\ref{I6}) and
\begin{equation}
d\ll \delta_0=c/\omega_p.
\label{N2}
\end{equation}
Dots in Eq.~(\ref{2}) denote terms of higher order of the small parameter
$d/\delta_0$. The procedure (\ref{6}) of the regularization of the Green
function allows us to avoid the "surface" divergence in the Casimir force.
The divergent term does not depend on the separation $a$ between the plates
and represents an addition to the renormalized Casimir force. According to
Eqs.~(\ref{3}) and (\ref{7}), the expression for this force
is
\begin{equation}
f=-kT\int\limits_0^\infty\frac{{\rm d}q^2}{2\pi}
{\sum_{n=0}^\infty}^\prime\frac{\epsilon k^2d\exp(-2ka)}
{2+\epsilon kd[1-\exp(-2ka)]}.
\label{8}
\end{equation}

For Casimir's interaction of sufficiently thin films, the characteristic
frequencies of the fluctuating electromagnetic field turn out to be much less
than the parameter $c/a$,
\begin{equation}
\omega_c\ll c/a.
\label{N3}
\end{equation}
In this case, one can neglect the relativistic retarding effect and passage
to the limit $c\to\infty$~\cite{19}. This allows us to assume $k=q$
in Eq.~(\ref{8}) for the characteristic frequencies and to approximate the
Casimir force as
\begin{equation}
f=-\frac{B}{4\pi\beta a^3}\int\limits_0^\infty {\rm d}x x^3
I(x)e^{-x},\qquad \beta =\frac{1}{kT}
\label{9}
\end{equation}
where $x=2qa$ is the new variable of integration, symbol $I(x)$ denotes the
sum of the series,
\begin{equation}
I(x)={\sum_{n=0}^\infty}^\prime\frac{1}{n(n+C)+BF(x)},
\label{10}
\end{equation}
with the parameters
\[
B=\frac{\omega_p^2\beta^2}{(4\pi )^2}\frac{d}{a}, \qquad
C=\frac{\beta\nu}{2\pi}
\]
and the function $F(x)=x(1-e^{-x})$.

In the case (\ref{N3}), the temperature-dependent part
of the Casimir force can be calculated by means of Eq.~(\ref{9}) without
using summation formula (\ref{A4}) for the Casimir force. However, the
analysis of the spectral integrals in (\ref{A4}) is useful in order to
define the characteristic frequencies giving the main contribution to the
Casimir force $f$. Let us recall that the contribution $f_0$ of the vacuum
fluctuating electromagnetic field to the Casimir force is defined by the
spectral density of energy taken at the imaginary frequency whereas the
contribution $\Delta f(T)$ of the thermal radiation of the system is related
to the imaginary part of the spectral density at the real frequencies. The
force $f_0$ connected to the vacuum fluctuations is evaluated by
Eq.~(\ref{I5}). It disappears with the decrease of the film thickness,
$d\to 0$.

Now consider the temperature-dependent part of the Casimir force.
According to Eqs.~(\ref{A4}) and (\ref{9}), we have
\begin{equation}
\fl
\Delta f(T)=-\frac{B'}{4\pi a^3}\int\limits_0^\infty {\rm d}x x^3
e^{-x}\int\limits_0^\infty {\rm d}\tau (e^{2\pi\tau}-1)^{-1}
\frac{\tau}{(\tau^2 -BF(x))^2 +C^2\tau^2},
\label{11}
\end{equation}
where $\tau =2\pi\beta\omega$. At $'\to 0$, this expression can be
approximated as
\begin{eqnarray}
\Delta f(T)&=-\frac{B}{4 a^3}\int\limits_0^\infty {\rm d}x x^3
e^{-x}\int\limits_0^\infty {\rm d}\tau (e^{2\pi\tau}-1)^{-1}
\delta [\tau^2 -BF(x)]\nonumber\\
&=-\frac{\omega_c\beta}{32\pi a^3}\int\limits_0^\infty
{\rm d}x x^3 e^{-x}F^{-\frac{1}{2}}(x)\left(e^{\frac{1}{2}\beta\omega_c
\sqrt{F(x)}}-1\right)^{-1}.
\label{11a}
\end{eqnarray}
The corresponding characteristic frequency $\omega_c =\omega_p\sqrt{d/a}$ of
the Casimir interaction is less than the parameter $kT/\hbar$
($\beta\omega_c\ll 1$) for sufficiently small thicknesses $d$. In this case
\begin{equation}
\Delta f(T)=-1.2\frac{kT}{8\pi a^3}-f_0+\cdots ,
\label{11b}
\end{equation}
where the symbol $\cdots$ denote terms of higher order of the smallness.
Thus,
\begin{equation}
f(T)=f_0 + \Delta f(T)=-1.2\frac{kT}{8\pi a^3}+\cdots .
\label{12}
\end{equation}
Asymptotics (\ref{12}) can be shown to be valid at $B,C \ll 1$ as well as
for $B,1 \ll C$. In terms of the characteristic frequencies these
inequalities give
\begin{equation}
\omega_c\ll{\rm max}\left(\frac{kT}{\hbar},\sqrt{\nu\frac{kT}{\hbar}}\,\,
\right).
\label{HT}
\end{equation}
If the inverse inequality is fulfilled we get evaluation (\ref{I7}).

The surprising things are that formula (\ref{12}) coincides with
Eq.~(\ref{I5}) for the Casimir force of the bulk metals and that $f(T)$ does
not disappear even at $d\to 0$. The difference between the cases of the bulk
materials and the thin films consists only in the inequality defining the
high-temperature regime.  This is the specific feature of the Casimir
attraction of the {\it metals}.  Contrary to the dielectrics, a significant
reduction of the characteristic frequencies of the thermal fluctuations in a
metal film occurs if the film thickness decreases. In its turn, this
leads to the growth of the effective dielectric permittivity (\ref{epsil})
of the conductor and, as it follows from Eq.~(\ref{7}), to a disappearance
of the dependence of temperature Green's function ${\rm reg}\Gamma$ on the
sample thickness. It is precisely this fact that results eventually in the
unexpected insensibility of the Casimir force to the thickness $d$ in the
high-temperature regime (\ref{HT}).

Using the asymptotics (\ref{I7}) and (\ref{12}), we can obtain the following
evaluating formula for the Casimir force:
\begin{equation}
f\propto -\left(kT+\frac{\hbar\omega_c^2}{\nu
+\omega_c}\right)\frac{1}{a^3}.
\label{13}
\end{equation}
It is necessary to keep in mind that Eq.~(\ref{13}) is obtained under
conditions~(\ref{I6}) and (\ref{N3}) for the frequency $\omega_c$
and for sufficiently thin films with $d$ satisfying the inequality
(\ref{N2}).

\section{Discussion}
The theoretical description of the temperature dependence of the Casimir
force between a bulk conductor and a thin metal film is given in the present
paper. In the general case, the Casimir force can be presented as a sum
(\ref{A4}) of temperature-dependent and a vacuum contributions of
fluctuating electromagnetic field. We have obtained the surprising result
for the situation of reasonably thin films (or reasonably high temperatures)
Eq.~(\ref{HT}). The Casimir force Eq.~(\ref{12}) has proved to be independent
of the sample thickness in the main approximation. This fact is
characteristic precisely for the metals because the Casimir force for
dielectric films with a constant value of $\epsilon$ vanishes at $d \to 0$.

Mathematically, the mentioned above $d$-independence of the Casimir
attraction of the metal film in the regime (\ref{HT}) is connected to the
proportionality of the characteristic frequency $\omega_c$ of the thermal
fluctuations to the thickness of the film and to the significant reduction
of $\omega_c$ with the decrease of $d$. The appearance of the characteristic
low-frequency regime in the Casimir attraction of the sufficiently thin
metal films is physically caused by the strong classical long wavelength
fluctuations of the conduction current and the plasmic shielding of the
electromagnetic modes.  These peculiarities add to the list of the
characteristic features \cite{13,14} of the Casimir effect for metals.

In conclusion, let us emphasize the entropy origin of the Casimir force
(\ref{12}). As it follows from (\ref{12}), the free energy of the Casimir
interaction has a form
\begin{equation}
{\cal F}=-1.2\frac{kT}{16\pi a^2}A.
\label{14}
\end{equation}
Hence, the entropy of interaction is
\begin{equation}
S=1.2\frac{A}{16\pi a^2}.
\label{15}
\end{equation}
Contrary to the free energy and entropy, the energy $E$ of interaction
calculated by the formula,
\begin{equation}
E=\frac{\partial}{\partial\beta}(\beta{\cal F}),
\label{16}
\end{equation}
vanishes at $d\to 0$.

\appendix
\section{Summation formula for the Casimir force}
We transform the sum Eq.~(\ref{1}) over Matsubar's frequencies $\omega_n$
into the "spectral" integrals using the Abel-Plan formula for
summing up a series,
\begin{eqnarray}
\fl
\lim_{n\rightarrow\infty}\left\{\sum_{k=1}^n
F(k)-\int\limits_{\theta}^ {n+\theta}{\rm
d}xF(x)\right\}&=\int\limits_{\theta}^{\theta-{\rm i}\infty} {\rm
d}zF(z)\left(e^{2{\rm i}\pi z}-1\right)^{-1} \nonumber\\
&+\int\limits_{\theta}^{\theta+{\rm i}
\infty}{\rm d}zF(z)\left(e^{-2{\rm i}\pi z}-1\right)^{-1},
\label{A1}
\end{eqnarray}
where the function $F(z)$ is regular in the half-plane ${\rm Re}z>0$ and
satisfy the inequality
\[
\vert F(x+{\rm i}y)\vert<f(x) e^{a\vert y\vert} \quad (a<2\pi);
\]
the function $f(x)$ is bounded at $x\to \infty$; $0<\theta<1$.
The applicability of the Abel-Plan formula (\ref{A1}) to the expression for
Casimir's force (\ref{1}) is ensured by the analytical properties of the
conductivity $\sigma^{(M)}$ and of the Green function $\Gamma^{(M)}$ in the
upper half-plane of the complex frequency~\cite{15,16}. Then, we make the
limiting transition $\theta\rightarrow 0^+$ in Eq.~(\ref{A1}) ($0^+$ is an
infinitesimal positive parameter). The final transformations are carried out
with regard to the formulae

\begin{equation}
\Gamma^{(M)}(\omega_n)=\Gamma({\rm i}|\omega_n|), \qquad
\Gamma(\omega)=\Gamma^{(M)}\left(\frac{\omega}{\rm i} + 0^+\right)
\label{A2}
\end{equation}
where $\Gamma$ is the retarding Green function of the electromagnetic
field, and the symmetry relations
\begin{equation}
\sigma^*(-\omega) =\sigma (\omega), \qquad
\Gamma^*(-\omega) =\Gamma(\omega).
\label{A3}
\end{equation}
As a result, we get Eq.~(\ref{A4}) for the Casimir force expressed in terms
of the spectral integrals.

{}

\end{document}